\documentclass[sigconf]{acmart}
\usepackage{hyperref}
\usepackage{enumitem}
\usepackage{float}
\usepackage{array}
\usepackage{booktabs}
\usepackage{tabularx}
\usepackage[normalem]{ulem}
\usepackage[most]{tcolorbox}
\usepackage{xcolor}
\usepackage{graphicx}

\usepackage{geometry}
\tcbuselibrary{skins,breakable}

\definecolor{userbg}{HTML}{DCEEFB}
\definecolor{userborder}{HTML}{1F6FB2}
\definecolor{aibg}{HTML}{F2F2F2}
\definecolor{aiborder}{HTML}{555555}
\definecolor{catbar}{HTML}{2E7D32}
\definecolor{intentchip}{HTML}{1F6FB2}
 
\AtBeginDocument{%
  }

\setcopyright{none}
\renewcommand\footnotetextcopyrightpermission[1]{}

\acmConference[Preprint]{UK}{August}{2026}
\begin{document}

%%
%% The "title" command has an optional parameter,
%% allowing the author to define a "short title" to be used in page headers.
%\title{[Placeholder] Understanding how Financial Services are discussed in Conversational AI across US and India}
\title{From Information to Delegation: Mapping Human-AI Financial Decision Making}

%%
%% The "author" command and its associated commands are used to define
%% the authors and their affiliations.
%% Of note is the shared affiliation of the first two authors, and the
%% "authornote" and "authornotemark" commands
%% used to denote shared contribution to the research.
\author{Iman Munire Bilal}
\email{iman.bilal@stripepartners.com}
%\orcid{1234-5678-9012}
\affiliation{%
  \institution{Stripe Partners}
   \city{}
  \country{}
}

\author{Yingcan Carol Wang}
\email{carol.wang@stripepartners.com}
%\orcid{1234-5678-9012}
\affiliation{%
  \institution{Stripe Partners}
  \city{}
  \country{}
}

\author{Ajan Raj}
\email{ajan.raj@stripepartners.com}
%\orcid{1234-5678-9012}
\affiliation{%
  \institution{Stripe Partners}
   \city{}
  \country{}
}

\author{Filippo Giovagnini}
\email{filippo.giovagnini@stripepartners.com}
%\orcid{1234-5678-9012}
\affiliation{%
  \institution{Stripe Partners, Imperial College}
   \city{}
  \country{}
}

\author{Pranav Tewari}
\email{pranav.tewari@stripepartners.com}
%\orcid{1234-5678-9012}
\affiliation{%
  \institution{Stripe Partners}
   \city{}
  \country{}
}

\author{Yuwei Zhang}
\email{yuwei.zhang@stripepartners.com}
%\orcid{1234-5678-9012}
\affiliation{%
  \institution{Stripe Partners}
   \city{}
  \country{}
}

\author{Mei-Chen Zoe Liou}
\email{zoe.liou@stripepartners.com}
%\orcid{1234-5678-9012}
\affiliation{%
  \institution{Stripe Partners}
   \city{}
  \country{}
}

\author{Qamar Zaman}
\email{qamar.zaman@stripepartners.com}
%\orcid{1234-5678-9012}
\affiliation{%
  \institution{Stripe Partners}
  \city{}
  \country{}
}

%%
%% By default, the full list of authors will be used in the page
%% headers. Often, this list is too long, and will overlap
%% other information printed in the page headers. This command allows
%% the author to define a more concise list
%% of authors' names for this purpose.
\renewcommand{\shortauthors}{Bilal et al.}

%%
%% The abstract is a short summary of the work to be presented in the
%% article.
\begin{abstract}
%This study investigates the role of conversational AI in financial decision-making across 1.5 million real-world ChatGPT and Gemini interactions from users in the United States and India. We propose a behavioral framework that classifies interactions by intent and delegated authority to measure how much and in which financial contexts, agentic assistance is occurring. The results indicate that while financial services constitute a significant use case for conversational AI, users primarily leverage these tools to inform and shape financial judgment rather than to delegate execution. By establishing this behavioral baseline, this work provides a foundation for assessing the transition of LLMs from information-retrieval tools to decision-making partners in financial ecosystems.
As AI increasingly participates in human decision making, understanding how decision-making authority is distributed between humans and AI has become a fundamental behavioural question. We introduce a behavioural measurement framework combining intent and delegated decision authority to quantify what consumers seek from AI and how much decision-making authority they assign to it. Applied to 1.5 million real-world ChatGPT and Gemini interactions from 6,304 users in the United States and India, we find that financial services are already a substantial AI use case. Consumers overwhelmingly use AI to retrieve information and shape financial judgement, while delegation of financial execution remains rare. By shifting attention from conversation topics to delegated decision authority, this work establishes a behavioural baseline for measuring the transition to increasingly agentic AI.

\end{abstract}


% \received{20 February 2007}
% \received[revised]{12 March 2009}
% \received[accepted]{5 June 2009}

%%
%% This command processes the author and affiliation and title
%% information and builds the first part of the formatted document.
\maketitle

\section{Introduction}

%\sout{Since the release of OpenAI's first ChatGPT model in 2022~\cite{openai2022chatgpt},} 
Large language models are rapidly evolving from systems that answer questions into systems that actively support human decision-making, with profound implications for the global economy \cite{google_ai_economy}. %\sout{By July 2025, ChatGPT was used by approximately 10\% of the world's population~\cite{chatterji2025how}.} 
This transformation is particularly significant in financial services, where decisions are characterised by uncertainty, complex trade-offs, regulatory constraints, and potentially irreversible consequences. As conversational AI moves from an information retrieval tool to a decision-making partner, understanding how decision authority is allocated between humans and AI becomes a fundamental behavioural question.

\begin{figure}
    \centering
    \includegraphics[width=0.4\textwidth]{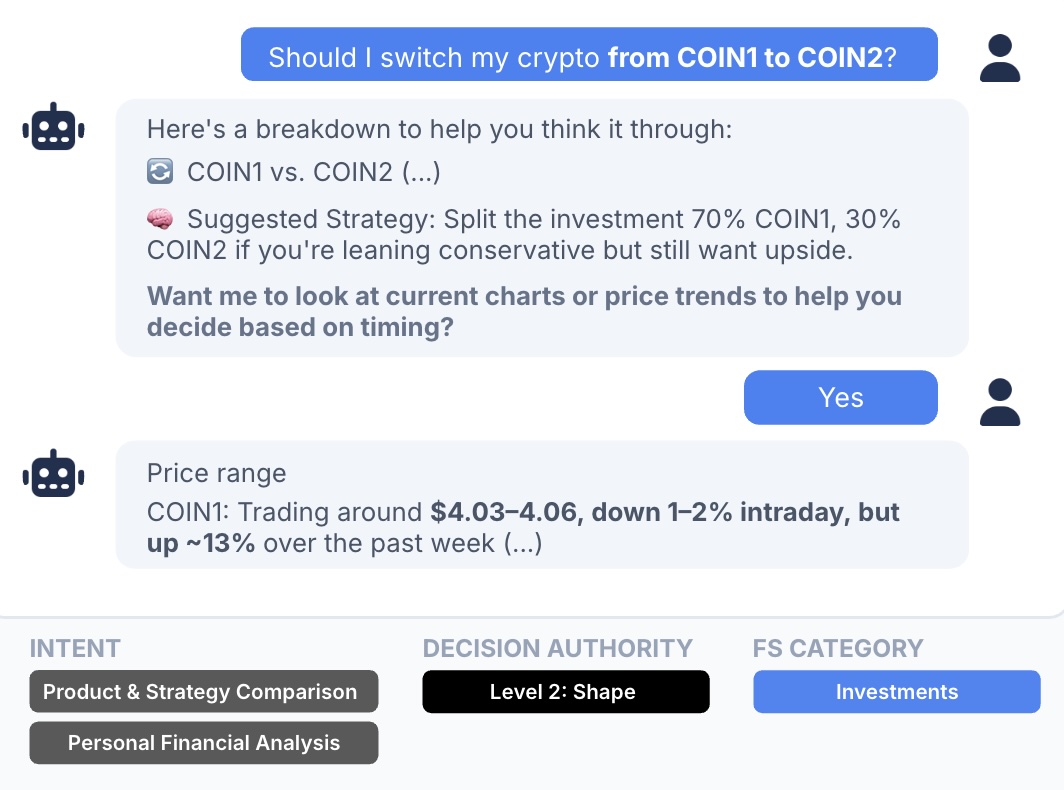}
    \caption{Example of conversation classified by our behavioral framework for behavioural intent, allocated decision authority and financial services category.}
\end{figure}

Recent research has substantially advanced our understanding of conversational AI through analyses of general usage patterns, occupational applications, and high-stakes domains such as healthcare~\cite{shelby2025taxonomyuserneedsactions,he2023llmhealthcare}. Complementary work has examined AI-generated financial advice and evaluated the financial capabilities of LLMs~\cite{nie2024llmfinance}. However, these studies largely characterise financial conversations by their topics or evaluate AI outputs in isolation, rather than examining how AI is incorporated into the broader process of financial judgement and decision making.

We argue that financial conversations should not be merely understood as collections of topics, but as behavioural episodes within a broader process of human–AI decision making. We introduce a behavioural framework that combines user intent and levels of delegated decision authority within financial domains, enabling conversations to be characterised by both the purpose they serve and the role AI plays within them.

We propose that the central question is not simply what people ask AI, but how much decision authority they assign to it. To capture this, we introduce decision authority as a behavioural lens for understanding human–AI financial interactions. Rather than treating conversations solely as topical exchanges, decision authority characterises whether AI is used to inform decisions, shape them, or act on behalf of users. This perspective provides a way to quantify how cognitive and decision-making responsibility is increasingly distributed between humans and AI.

Applying this framework to more than 1.5 million real-world ChatGPT and Gemini interactions of users in the US and India reveals that financial services already constitute one of the largest application domains for conversational AI with approximately half of users engaging in finance conversations during the study period. In both countries, consumers mainly use AI to inform and shape financial decisions, while delegation of financial execution remains rare and is largely confined to budgeting and financial tracking.

Our work makes two contributions. First, we introduce a measurement framework for understanding human–AI financial interactions that combines two complementary dimensions: behavioural intent, which captures what users are trying to accomplish, and decision authority, which captures the level of decision-making authority they allocate to AI. Second, we provide the first large-scale empirical characterisation of how consumers incorporate conversational AI across the financial decision-making process, establishing a behavioural baseline against which increasingly agentic AI systems can be evaluated.

\section{Related Work}

Several recent studies have analysed large collections of human--AI conversations to characterize common usage patterns. Chatterji et al.~\cite{chatterji2025how} present one of the first large-scale analyses of ChatGPT conversations, introducing a taxonomy of user intents and tasks across multiple domains. The Anthropic Economic Index~\cite{handa2025anthropic} complements this perspective by analysing millions of Claude interactions to characterize the distribution of AI-assisted work across occupations and economic activities. Similar large-scale analyses have been conducted using Microsoft Copilot conversations~\cite{costagomes2025time,costagomes2026health} and exported ChatGPT user chat histories across multiple countries~\cite{roychowdhury2026conversation}. %Collectively, these studies provide a broad understanding of how conversational AI is used in practice.

Recent work has shifted to LLMs in financial decision-making. Paydarzarnaghi et al.~\cite{paydarzarnaghi2026financechatgpt} analyse real-world ChatGPT conversations to identify financial topics discussed with AI, and Pak~\cite{pak2026personalfinance} examines how generative AI supports everyday personal finance tasks such as budgeting, investing, and planning. Complementary work evaluates the quality and impact of AI-generated financial advice: Choukhmane et al.~\cite{choukhmane2025financialadvice} study demand for AI-assisted advice and its effects on household decisions, while Niszczota and Abbas~\cite{niszczota2023financially} assess GPT ’s literacy and potential as financial advisor.

A related line of work models financial intents and user behavior. BANKING77 ~\cite{casanueva2020banking77} is a standard dataset for banking intent classification, while industry analysis from Lloyds Banking ~\cite{lloyds2025digitalindex} shows how consumers use AI for banking and financial information-seeking.

Despite progress, prior work has mainly focused on high-level patterns of conversational AI use or the quality of LLM-generated financial advice. We instead examine the behavioural role of AI in financial decisions. We propose a measurement framework that combines financial domains, behavioural intent, and decision authority to capture not only what users ask AI, but what they aim to achieve and how much authority they delegate to AI. We apply this framework to large-scale real-world conversations from the US and India to characterise emerging patterns of human–AI financial interaction.

\section{Data}

\subsection{Source}

\begin{table}[H]
\centering
\small
  \caption{Overview of full dataset size with respect to users and user messages across US/India and ChatGPT/Gemini.}
  \label{tab:dataset_overall_stats}
  \begin{tabular}{lccc}
    \toprule
    &US&India & Total\\
    \midrule
    \# of users & 2499 & 3805 & 6304\\
    \# of user prompts & 760.3k & 766.1k & 1.53M\\
    \# of ChatGPT user prompts & 691.3k & 620.5k & 1.31M\\
    \# of Gemini user prompts & 69k & 145.6k & 214.6k\\
  \bottomrule
\end{tabular}
\end{table}
\normalsize

The corpus covers the ChatGPT and Gemini histories of conversations during August - October 2025 from a sample of 2499 US-based users and 3805 India-based users (See Table \ref{tab:dataset_overall_stats}) recruited and reimbursed by MeasureProtocol data provider. The sample recruitment was done on a voluntary opt-in basis. 

The corpus contains metadata to indicate  user messages, AI responses and timestamps of each. %\sout{For ChatGPT, conversation histories are further divided into chats, while for Gemini, there is no access to this feature.}

\subsection{Sample Representation}

We benchmark the representativeness of our studied user sample against CENSUS available data of the general adult population in US and India with respect to age, gender, income and employment status 
\footnote{\url{https://www.census.gov},
\url{https://population.un.org/wpp}, \url{https://www.mospi.gov.in}}. Although we compare our sample with GenPop, the more appropriate benchmark is active conversational AI users, who are known to be younger and demographically distinct. Some deviation from population benchmarks is therefore expected rather than indicative of sampling bias.

In the US, the sample skews female (58.1\% of sample vs 51.0\% GenPop are women), younger (58.2\% of sample vs 29.0\% GenPop are aged 18-34), and lower-income (52.0\% of sample vs 30.2\% of GenPop are in under \$50k income households)
%\sout{. With respect to employment status, employment rates are higher in the study sample (70.1\% of sample users vs 61.0\% GenPop are employed)}, 
with fewer respondents in unemployment and inactive labour roles (e.g., retirees, homemakers) compared to Census benchmarks.

The Indian sample is heavily male (76.5\% of sample vs 51.3\% GenPop are men) and younger (89.4\% of sample vs 41.2\% GenPop are aged 18-34) with lower employment rates (48.1\% of sample vs 57.4\% GenPop are employed). Household comparison is not possible due to lack of recently available income data for GenPop India.

\section{Methodology}
%The methodology section outlines: (a) the data processing and cleaning of the full Human-AI corpus and (b) the NLP modelling of financial aspects discussed in the corpus.  (a) involves a translation and chat segmentation step; (b) describes the identification of the finance conversations in the corpus, financial product tagging, intent classification and topic modelling.

The methodology section outlines: (a) the data processing and cleaning of the full Human-AI corpus which involves a translation and chat segmentation step and (b) the NLP modelling which describes the identification of the finance conversations in the corpus, financial product tagging, intent classification and topic modelling.

\subsection{Data Processing}
\subsubsection{Translation}
We translate all non-English conversations from the India segment into English using Google Translate packages – this step is relevant to 144k non-English messages. This processing step compensates for the paucity in data resources and models for non-Western languages \cite{Joshi2020State} and allows better comparability of results between markets in the modelling stage.

\subsubsection{Chat Segmentation}
For both AI sources, we observe that the raw Human-AI conversations often contain multiple unrelated topics that are consecutively researched with the AI assistant within a single session \footnote{We hypothesize this is happening due to users not initiating a new "chat" with the AI assistant for each new topic of discussion and instead continue conversing in the existing session.} – this is similar to the 'topic drift' phenomenon occurring in long online discussion threads \cite{Park2016TopicDrift}. %\sout{This behaviour is further evidenced by the long time span between the first and last user message in a session spanning more than 10 user prompts: ChatGPT entries take on average 1.1 days and Gemini ones 79.7 days.}

In order to standardise the conversations between the two platforms, all raw user histories in our dataset were topically segmented into smaller units coined 'subchats' where each subchat covers only one topic. For the segmentation process, we impose the rules:

\begin{itemize}
    \item Only attempt segmentation for sessions longer than 10 user prompts; shorter sessions  will remain as is and form 1 subchat.

    \item Within a long session, if two consecutive user prompts share low-frequency terms in common (high-frequency terms or stopwords do not count), these prompts and their AI responses should be appended to the same subchat. 
    
    \item If a user prompt is too short (under 3 words), the user prompt and its associated AI response should be appended to the existing subchat. 

    \item Within a session, consecutive user prompts with no overlap of low-frequency terms and longer than 3 words will be assigned to different subchats.

\end{itemize}

The above heuristics have been manually evaluated on a small test set of Human-AI conversation histories before being applied to the full dataset. This process results in 291.9k subchats for US and 213.4k subchats for India.

\subsection{Modelling}

\subsubsection{Finance subset identification}
\label{subsection:finance_subset_identification}

For the context of generalist conversational AI, we broadly define \emph{finance} as any discussion where the user engages with the LLM about financial services, income-generating methods or other money-related issues. 

\noindent \textbf{Task definition} This process is modelled as a binary detection task $f:D \rightarrow \{0,1\}$, where $D$ is the set of all subchats in our corpus, such that for each subchat $s\in D$:
\begin{equation}
 f(s) =
  \begin{cases}
  1 & \text{if } s \text{ discusses finance} \\
  0 & \text{otherwise}
  \end{cases}
\end{equation}

Note that we construct $f$ via a fine-tuned Transformer-based classification model ~\cite{vaswani2017} to capture the latent semantic representation of finance discussions.

\noindent \textbf{Training dataset} This comprises 5.8k subchats sampled so they include conversations from both markets and both AI platforms to account for language drift differences. To ensure we capture candidates for finance conversations in the golden dataset, we first curate a dictionary of 2.4k financial keyword and sample conversations which mention at least one of these. To account for non-finance subchats, we sample subchats that do not mention these keywords. The final dataset was annotated by 3 members of the research team.

\noindent \textbf{Model} We choose a LongFormer model ~\cite{Beltagy2020Longformer} in favour of BERT-like candidates ~\cite{devlin-etal-2019-bert} as the former has a much longer context window (4096 tokens) more suitable to long documents (the average subchat length is 2047 tokens). This is obtained by enabling the attention pattern to scale linearly instead of quadratically, like for BERT-like architectures, via a sparsified self-attention matrix.
%{\color{red} FG: This is obtained by making the attention pattern scale linearly instead of quadratically, like for BERT-like architectures. This is attained by sparsifying the full self-attention matrix. For more details, see \cite[Figure 2, Section 3]{Beltagy2020Longformer}}. 
The pre-trained model is further fine-tuned on a manually annotated dataset extracted from the conversation corpus.

\noindent \textbf{Setup \& Evaluation} The model is fine-tuned on 80\% on the annotated dataset and tested on the remaining 20\%. The training is conducted for 4 epochs with learning rate = 2e-5, batch size = 4. The best checkpoint with respect to accuracy on test set is saved and used for inference on the full dataset – its performance achieves $Accuracy = 96.5$ and $F_1 = 97.3$.

Model inference is conducted on the full dataset $D$ and any subchats predicted as finance by the trained classifier above will be denoted $D_{finance}$.

\subsubsection{Tagging of Financial Services and Products}
\label{subsection:fs_service_product_tagging}

\begin{table}
    \centering
    \small
    \caption{Schema for tagging conversations by Financial Services and Products. Starred categories are in work of ~\cite{theerthala-2025-synthesizing}.}
    \label{tab:fs_cat_subcat_schema}
    \begin{tabular}{p{1.5cm}p{3.5cm}p{2.5cm}}
    \toprule
    \textbf{Category} & \textbf{Subcategories} & \textbf{Definition}\\
    \midrule
Investments &
\emph{Liquid Securities}
\newline \emph{Alternative Assets}
\newline \emph{Account Types}
\newline \emph{Strategy \& Analysis} & Growth-oriented assets and capital market participation \\
\midrule
 Retail Banking \& Credit &
\emph{Deposit Products}
\newline \emph{Asset-Backed Lending}
\newline \emph{Unsecured Lending}
\newline \emph{Instruments \& Monitoring} & Individual liquidity and consumer debt \\
\midrule

Tax &
\emph{Income \& Output Taxes}
\newline \emph{Asset-Based Taxes}
\newline \emph{Tax Accounting}
 & Statutory obligations and government levies \\
\midrule

Payments \& Transfers$^{*}$ &
\emph{Transfer Services}
\newline \emph{Payment Infrastructure}
\newline \emph{Stored Value}
 & The movement of value and transaction processing \\

\midrule
Benefits \& Public Aid$^{*}$ &
\emph{Direct Assistance}
\newline \emph{Healthcare Support}
\newline \emph{Disability \& Retirement Support}
\newline \emph{Education Support}
 & Non-market financial support and social safety nets \\

\midrule
Insurance &
\emph{Life \& Health}
\newline \emph{Liability \& Property}
\newline \emph{Policy Mechanics}
 & Contingent contracts for risk transfer \\

  \midrule
Business Finance$^{*}$ &
\emph{Commercial Credit}
\newline \emph{Operational Finance}
\newline \emph{Strategic Finance}
 & Corporate and entity-level financial management \\

  \midrule
Finance Infrastructure$^{*}$ &
\emph{Regulatory Bodies}
\newline \emph{Verification \& Identity}
\newline \emph{General Infrastructure}
 & Financial ecosystem infrastructure and compliance \\

  \midrule
Budgeting &
\emph{Tracking}
\newline \emph{Planning}
 & Behavioral planning and cash flow management \\
         \bottomrule
    \end{tabular}
\end{table}

\normalsize

Building on previous work exploring how personal finances are researched on and influenced by social media channels ~\cite{WARKULAT2024103721,cao2020}, we shift our focus to understand what products and services relevant to the financial sector (FS) are prevalent in user conversations with LLMs. To capture this, we define a MECE framework of financial categories and associated subcategories – this is inspired by the categorisation of financial user queries by ~\cite{theerthala-2025-synthesizing} and further augmented by empirical evidence of other categories present in our finance dataset. The framework is shown in Table \ref{tab:fs_cat_subcat_schema}.

\noindent \textbf{Task formulation} The task is two-staged: 
\begin{enumerate}
    \item \emph{Entity extraction}: Similar to the extraction setup by ~\cite{lu-huo-2025-financial}, we use an LLM to extract all entities relevant to finance services and products mentioned in subchats $D_{finance}$ and standardise entities discussing the same concept (e.g., “S\&P 500”, ”VOO“, ”SPY” all indicate indexes of the "S\&P 500") to minimise duplication. The final keyword dictionary is denoted $K$. Each keyword in $K$ is then manually mapped to its corresponding categories and subcategories in the FS framework or when none apply, it is  removed from the keyword dictionary (See Evaluation step). Formally this is defined as $f:K \rightarrow 2^{N_{cat}} \times 2^{N_{subcat}}$ where $N_{cat}$ and $N_{subcat}$ are the sets of categories and subcategories and for each keyword $k\in K$:
    $$f(k)= (n_{cat},n_{subcat})$$

    where $n_{cat}\subseteq N_{cat}$ and $n_{subcat}\subseteq N_{subcat}$. Note that our notation allows for some keywords in the curated dictionary $K$ to be mapped to multiple categories when relevant.
    
    \item \emph{Taxonomy matching}: Using the manually-created mapping created above, the process to tag subchats by FS categories is deterministic. Namely, for each subchat $s \in D_{finance}$, we assign a corresponding category and subcategory if $s$ mentions at least one keyword relevant to these (e.g., a subchat mentioning "S\&P 500" is associated to category \emph{Investments} and subcategory \emph{Liquid Securities}). 
\end{enumerate}

\noindent \textbf{Model} We employ \emph{GPT 4o-mini}\footnote{\url{https://developers.openai.com/api/docs/models/gpt-4o-mini}} for the initial extraction of FS terms in subchats as it balances low computing costs when applied at scale against single-task performance when prompt-engineered. 

\noindent \textbf{Evaluation} The process for entity extraction employs an LLM step to detect which FS entities are discussed in each conversation in the finance dataset. This yields a ranking of most mentioned keywords, allowing the evaluation to focus on removing frequently-occurring false positives erroneously introduced by the LLM. The process is validated by one member of the team who reviews each frequent keyword's fit against the FS framework (\ref{tab:fs_cat_subcat_schema}). This leads to 2.7k correctly matched FS keywords and 163k unique unmatched keywords. A review of the top 25 most mentioned unmatched terms reveals these are general and uninformative terms (raw monetary amounts, generic money words) or of small importance for financial services (reward apps, gig platforms, betting-related terms).

All subchats in the finance dataset $D_{finance}$ mentioning at least one FS keyword will be denoted $D_{FS}$.

\begin{table*}
\centering
\small
  \caption{Taxonomy of user intents in financial Human-AI conversations with definition, examples and decision authority level.}
  \label{tab:intent_taxonomy}
  \begin{tabular}{lp{5cm}p{4.5cm}p{2cm}}
    \toprule
    \textbf{Intent} & \textbf{Definition} & \textbf{Examples} & \textbf{DA level}\\
    \midrule
    Delegated Financial Decision Execution & Completing a financial transaction or commitment with some degree of decision freedom delegated to LLM & \textit{“Find the best savings account and move my money”\newline “Invest my money in the best option”} & 3\\
    \midrule
    Instruction-led Financial Execution & Completing a financial transaction or commitment with very defined parameters and instructions & \textit{“Transfer £500 to John via PayPal”\newline “Cancel my Ocado subscription”} & 3\\
\midrule
    Financial Automation \& Monitoring & Setting recurring rules, alerts, or optimisation over time & \textit{“Alert me if spending spikes” \newline "Auto-invest monthly"} & 3\\
\midrule
    Product \& Strategy Optimisation & Optimising the recommendation of financial products and strategies for the user & \textit{“Best mortgage for me?”\newline “Should I refinance?”} & 2\\
\midrule
    Product \& Strategy Comparison & Comparing two or more financial products or strategies mentioned by the user & \textit{“Barclays vs Santander cashback”}& 2\\
\midrule
    Financial Problem Resolution & Resolving issues with accounts, fraud, payments, money-related issues & \textit{“I was charged twice” \newline “Is this transaction fraud?”}& 2\\
\midrule
    Personal Financial Analysis & Interpreting and calculating (often user-specific) financial data & \textit{“Where is my money going?”\newline “What would be the tax for this salary?”}& 2\\

\midrule
    Financial Planning & Structuring future financial behaviour with step-by-step planning or timelines & \textit{“Plan my retirement”\newline “How to pay off debt in 3 months?”}& 2\\
    
\midrule
    Complex Research  & Multi-step research that synthesises information across sources, products, or market conditions & \textit{“Research savings accounts for a higher-rate taxpayer in Chennai”}& 1\\
\midrule
    Simple Retrieval & Retrieving a specific, well-defined piece of financial information & \textit{“What is the ISA allowance this year?”\newline “What is the early repayment charge on my mortgage?”}& 1\\
\midrule
    Financial Learning \& Education & Building conceptual understanding of a financial topic, product, or term & \textit{“How does compound interest work?” \newline“What is the difference between a stocks and shares ISA and a cash ISA?”}& 1\\
\midrule
    Creation & Generating content (email, code, images) & \textit{“Python code to calculate the net income” \newline “Respond to credit email”}& N/A\\

    \bottomrule
  \end{tabular}
\end{table*}

\normalsize

\subsubsection{Financial Intent}

Understanding the intents users bring into LLMs reveals the purpose they are trying to accomplish behind their prompts~\cite{shah2025} irrespective of the specific topical context surrounding the query. This is especially relevant in a sensitive domain such as finance, where users may be reluctant to disclose underlying goals or decision-making
processes due to social desirability bias \cite{Krumpal2013}.

Our proposed content taxonomy is found in Table \ref{tab:intent_taxonomy}. This has been informed by a) existing user intent taxonomies prevalent in AI conversation platforms such as Bing Chat ~\cite{shah2025}, Claude ~\cite{handa2025anthropic} and ChatGPT ~\cite {shelby2025taxonomyuserneedsactions}, b) internal financial services expertise %\footnote{One senior member of the team has worked and led behavioural science projects in the field of financial services. Two members of the team hold economics degrees.} 
\footnote{One senior author has over 20 years' experience in financial regulation, behavioural science and financial services strategy, including leadership roles. % at the UK Financial Services Authority and Financial Services Culture Board. 
Three other members of the research team have strong experience contributing to financial services research}
and c) exploration of a random subset of our Finance dataset. To validate the intent guidelines, the team conducted a pilot study to stress-test the intent taxonomy on a subset of real-world conversations and to improve guideline clarity for the subsequent annotation step.

Additionally, we map each financial intent to a decision authority (DA) level depending on the extent of delegation assigned to the LLM by the user: 
\begin{itemize}[nosep]
    \item Level 3 (Act): interactions where the AI role is transactional or autonomous with the aim to execute and operationalise actions for the user
    \item Level 2 (Shape): interactions where the AI role is analytical or advisory with the aim to influence user's choice or strategy
    \item Level 1 (Inform): interactions where the AI role is informational with the aim to provide information or explanations to the user
\end{itemize}

\noindent \textbf{Task definition} This is modelled as a multi-label multi-class classification task as more than one user intent out of all 12 proposed intents can emerge in the span of a conversation (e.g., "What is the APY of this savings account?" followed by "How does it compare to other banks?"). The model is defined as $f:D_{FS} \rightarrow \{0,1\}^{12}$ such that for each subchat $s\in D_{FS}$:
\begin{equation}
 f (s)_i =
  \begin{cases}
  1 & \text{if } s \text{ is associated to intent}_i \\
  0 & \text{otherwise}
  \end{cases}
\end{equation}
where $f$ is modelled via a long-context Transformers architecture similar to the step before.

\noindent \textbf{Training dataset} We sample 2.8k conversations classified into the Finance vertical in the previous step and ensure these are distributed across different markets and LLM types. The dataset is further augmented with 600 synthetic examples generated by model \emph{GPT 5.5}\footnote{\url{https://developers.openai.com/api/docs/models/gpt-5.5}} (with reasoning effort set to "xhigh" to produce sample semantic diversity) to compensate for the lack of real-world representation of higher-agentic intents\footnote{At the time of data collection, the August-October snapshot of ChatGPT and Gemini conversations contained a very small volume (under 1\%) of conversations with intent for Delegated Financial Decision Execution, Instruction-led Financial Execution or Financial Automation \& Monitoring}. The intent distribution in the training dataset as well as the trade-off proportion between synthetic and organic subchats within each intent is outlined in Table \ref{tab:intent_training_stats}. Note that the scores do not sum to 1 as the setup is multi-label classification with each data point being associated to 2 intent labels on average.

The final training dataset contains 3.4k subchats which have been manually annotated by four of the co-authors responsible for devising and testing the intent taxonomy.

\begin{table}[H]
    \centering
    \small
    \caption{Distribution of intent-labelled subchats in training dataset (\textit{\% in training dataset}) and proportion of synthetic data used within each intent (\textit{Synthetic \%})}
    \label{tab:intent_training_stats}
    \begin{tabular}{lp{1.8cm}p{1.6cm}}
    \toprule
    \textbf{Intent} & \% \textbf{of training} & \textbf{Synthetic} \%\\
    & \textbf{dataset} & \\
    \midrule
         Delegated Fin. Decision Ex. & 10.34 & 96.01 \\
         Instruction-led Fin. Execution & 18.53 & 82.03 \\
         Fin. Automation \& Monitoring & 9.40 & 82.76 \\
         Product \& Strategy Optimisation  & 20.18 & 3.36 \\
         Product \& Strategy Comparison & 11.40 & 0.51 \\
         Fin. Problem Resolution  & 16.38 & 12.95 \\      
         Personal Fin. Analysis & 25.34 & 34.42 \\
         Complex Research  & 22.13 & 0.13 \\
         Simple Retrieval & 26.49 & 9.68 \\
         Fin. Learning \& Education & 12.08 & 0.00 \\
         Creation  & 16.47 & 23.97 \\
         \bottomrule
    \end{tabular}
\end{table}

\normalsize

\noindent \textbf{Model} We fine-tuned the BigBird ~\cite{bigbird} base model, a sparse-attention based Transformer capable of handling long-context similar to ~\cite{Beltagy2020Longformer}.

\noindent \textbf{Setup \& Evaluation} The model is fine-tuned on 90\% of the annotated dataset and tested on the remaining 10\%. Training is conducted for 8 epochs with learning rate = 1e-5, batch size = 4; the best checkpoint with respect to $F1_{Micro}$ on test set is saved. To boost performance, we apply label-specific probability thresholding due to the high intent imbalance in the training. This increases performance by 2.6 $F1_{Micro}$ pts (from $F1_{Micro} = 68.0$ to $F1_{Micro} = 70.6$)
%(see Table \ref{tab:intent_performance_eval}) 
and leads to a more stable performance across individual intents as suggested by the macro score (from $F1_{Macro}=67.9$ to $F1_{Macro}= 72.3$). 

Once trained, the intent classifier is applied to the $D_{FS}$. %of FS-labelled subchats.

% \begin{table}[h]
%     \centering
%     \small
%     \caption{Evaluation of multi-label intent model}
%     \label{tab:intent_performance_eval}
%     \begin{tabular}{lcc}
%     \toprule
%          Model & $F1_{Micro}$ & $F1_{Macro}$\\
%     \midrule
%          Fine-tuned BigBird & 68.02 & 67.92\\
%          Fined-tuned + Threshold tuning BigBird & 70.64 & 72.30 \\
%     \bottomrule
%     \end{tabular}

% \end{table}

\subsubsection{Topic Modelling}

To complement product and intent taxonomies we perform topic modeling. We apply BERTopic \cite{grootendorst2022bertopic} to summarised subchats in $D_{FS}$, pooling US and India conversations during model estimation to obtain a common topic space and subsequently reporting topic prevalence separately by market. 

\noindent \textbf{Embedding models}. To select the most suitable embedding, we consider the top models satisfying the trade-off between model performance and size as evaluated in the MTEB clustering leaderboard \cite{muennighoff2022mteb}, which at the time of the experiments were \href{https://huggingface.co/FinLang/finance-embeddings-investopedia}{FinLang/finance-embeddings-investopedia}, Qwen3-0.6B \cite{zhang2025qwen3embedding}, and \href{https://huggingface.co/microsoft/harrier-oss-v1-270m}{microsoft/harrier-oss-v1-270m}. All models are open-source with size $<1 \text{B}$ parameters.

\noindent \textbf{Evaluation}. The models are tested under the same dimensionality-reduction and clustering configuration. Cluster quality is evaluated via the silhouette coefficient \cite{rousseeuw1987silhouettes}, %Davies--Bouldin index \cite{davies1979cluster}, 
Calinski--Harabasz index \cite{calinski1974dendrite} and topic coherence \cite{roder2015exploring}. We select the \emph{Harrier} model which achieves the strongest overall cluster separation, with a silhouette coefficient of (0.62)
%Davies--Bouldin index of (0.468), 
and Calinski--Harabasz score of (40{,}033.3), while maintaining coherence comparable to the best-performing alternative.

% \noindent \textbf{Topic estimation}
% Subchat embeddings are reduced to five dimensions using UMAP with (15) nearest neighbours, zero minimum distance, cosine distance, and random seed (42). The reduced representations are clustered using HDBSCAN with a minimum cluster size of (25), maximum cluster size of (7{,}000), Euclidean distance, and excess-of-mass cluster selection. Subchats assigned to cluster (-1) are treated as noise.

% Topic representations are then obtained using class-based TF--IDF over a vocabulary of unigrams and bigrams occurring in at least two documents, after removing English stop words. Each topic is represented by its 20 highest-weighted terms, and document-level topic probabilities are estimated.

Topic representations are then obtained by summarising the top 100 documents, ranked by intra-topic probabilities. To prevent %repeated interactions from 
a small number of highly active users from dominating the results, we use Gini index thresholding to remove 9 highly skewed topics.

% \noindent \textbf{User-concentration filtering}
% To prevent repeated interactions from a small number of highly active users from dominating the results, we calculate the Gini coefficient of user-level subchat contributions within each topic. Topics with (G>0.70) are excluded from comparative prevalence analyses, removing nine highly concentrated topics. Topic prevalence is then calculated separately for the US and India using the retained topic assignments.

\section{Results}

%In this section, we summarise our findings on a) exploring to what extent the financial sector and its categories are discussed with LLMs (Section 5.1), b) measuring which intents and decision authority levels are offloaded to LLMs across finance categories (Section 5.2) and c) which specific topics emerge (Section 5.3). The results are reported separately for US and India to account for market behavioural differences.
This section describes how LLMs are incorporated into financial decision making in US and India. We first establish the prevalence of finance conversations, before examining where LLMs are used in FS domains, and what intents and decisions are allocated to these.

%\subsection{Finance and Financial services}
\subsection{Financial decision contexts}
\noindent \textit{Where is conversational AI being used in finance?}
\label{sec:finance_financial_services}
\begin{table}[H]
    \centering
    \small
     \caption{Engagement with FS in full Human-AI dataset.}
    \label{tab:fs_broad_stats}
    \begin{tabular}{ccccccc}
    \toprule
    \textbf{} & \multicolumn{2}{c}{\textbf{\% of subchats}} & \multicolumn{2}{c}{\textbf{\% of users}}\\
    \midrule
    & US & India & US & India \\
    \cline{2-5}
        All & 5.1 & 6.3 & 50.0 & 44.3 \\
        ChatGPT & 5.0 & 6.5 & 53.4  & 60.7 \\
        Gemini & 6.1 & 4.2 & 26.2  & 13.7 \\
        \bottomrule
    \end{tabular}
\end{table}
\normalsize

\normalsize
% \noindent \textbf{Overall volume} %\sout{Employing the finance classifier (see \ref{subsection:finance_subset_identification}) and the subsequent keyword filtering (see \ref{subsection:fs_service_product_tagging}),} 
% We first measure user engagement with LLMs on FS (Table \ref{tab:fs_broad_stats}): 50.0\% of all users engage in a financial services related conversation in the US and 44.3\% of all users do in India. %\sout{The actual volume of relevant conversation is smaller:}
% In terms of conversations, 5.1\% and 6.3\% of total subchats in the 3-month available history in the data concerns FS in US and India, respectively. %\sout{When broken down by LLM type,} 
% This aligns with how often other LLMs are employed for finance: 6.7\% of Claude conversations relate to Business \& Financial Operations as found by \cite{handa2025anthropic}.

\noindent Financial services already constitute a substantial use case for conversational AI in both markets (see Table \ref{tab:fs_broad_stats}). During the 3-month observation period, 50.0\% of users in the US and 44.3\% in India engaged in at least one FS-related conversation. FS accounted for 5.1\% of subchats in the US and 6.3\% in India, aligning with how often other AI agents such as Claude are used for finance \cite{handa2025anthropic}. These findings indicate that LLMs are already embedded in everyday financial decision making rather than representing a niche application.

\begin{table}[H]
    \centering
    \small
    \caption{Breakdown of each FS category in FS data by conversation and user volume mentioning the category}
    \label{tab:fs_cat_volume_stats}
    \begin{tabular}{p{3cm}p{0.75cm}p{0.75cm}p{0.75cm}p{0.75cm}}
    \toprule
         \textbf{Category} & \multicolumn{2}{c}{\textbf{\% of subchats}} & \multicolumn{2}{c}{\textbf{\% of users}}\\
         \midrule
    & US & India & US & India\\
    \cline{2-5}
    Retail Banking \& Credit & 33.6 &32.1& 33.7 &28.8\\
    Payments \& Transfers & 24.8 &34.1& 26.7 &28.9\\
    Investments & 24.6 & 34.8 & 23.5 & 26.9\\
    Benefits \& Public Aid & 13.8 &3.4& 19.2 &6.9\\
    Tax & 11.1 &14.9& 18.4 &19.1\\
    Insurance & 7.5 &4.2& 16.4 &8.9\\
    Business Finance & 6.6 &5.3& 11.4 &9.0\\
    Finance Infrastructure& 4.1 &9.7& 11.0 &15.0\\
    Budgeting & 3.6 &2.2& 9.6 &5.2\\

\bottomrule    
    \end{tabular}
\end{table}

Table \ref{tab:fs_cat_volume_stats} summarises the FS domains in which conversational AI is used. Across both countries, \emph{Retail Banking \& Credit}, \emph{Payments \& Transfers} and \emph{Investments} account for the majority of FS conversations, although cross-country differences emerge. In US, \emph{Retail Banking \& Credit} is the dominant category, with conversations concentrated on \emph{Credit Instruments \& Monitoring} and \emph{Deposit Products}. By contrast, Indian users devote a larger share of chats to \emph{Investments}, particularly concerning \emph{Liquid Securities} and \emph{Alternative Assets}. Investigation into the topics of these categories reveals \emph{Retail Banking \& Credit} conversations in US frequently concern improving credit scores, managing payments and choosing products, whereas Indian \emph{Investment} conversations are more focused on portfolio management, investment strategies and trading decisions. 

We also see that \emph{Insurance} accounts for a substantially larger share of FS interactions in US than India. While this study does not establish causal explanations, this difference is consistent with the greater maturity and product diversity of the US insurance market, creating more opportunities for consumers to seek information, compare products and navigate claims or coverage decisions.

\subsection{Behavioural Intent and Decision Authority}
\noindent \textit{What are consumers trying to accomplish, and how much authority do they assign to AI?}

\begin{table}[h]
    \centering
    \small
    \caption{Proportion of subchats and users engaging in FS conversations associated with each financial intent}
    \label{tab:fs_all_intent_stats}
    \begin{tabular}{lp{0.5cm}p{0.5cm}p{0.5cm}p{0.5cm}}
    \toprule
    \textbf{Intent \& Decision Authority} & \multicolumn{2}{c}{\textbf{\% of subchats}} & \multicolumn{2}{c}{\textbf{\% of users}}\\
    \midrule
    & US & India & US & India \\
    \cline{2-5}
        \textit{Act: Level 3 Decision authority} &0.3&0.1 &0.8&0.4\\ 
          Delegated Fin. Decision Execution &0.0&0.0 &0.0&0.0\\ 
          Instruction-led Fin. Execution &0.3&0.1&0.8&0.4 \\ 
          Financial Auto. \& Monitoring &0.0&0.0&0.2&0.0 \\ 
          \midrule
          \textit{Shape: Level 2 Decision authority} &58.6&49.7 &39.5&32.9\\
          Product \& Strat. Optimisation &14.4&17.7&22.6&19.3 \\ 
          Product \& Strat. Comparison &16.9&17.7&23.1&18.9 \\ 
          Financial Problem Resolution &27.8&19.7&24.4&20.2 \\ 
          Personal Financial Analysis &18.1&14.2&20.2&16.0 \\ 
          Financial Planning &6.4&7.6&12.8&12.5 \\
          \midrule
          \textit{Inform: Level 1 Decision authority} &63.5&72.1 &43.3&39.7\\
          Complex Research &21.2&26.3&30.0&25.0\\ 
          Simple Retrieval  &39.8&41.9&34.2&31.8 \\ 
          Financial Learning \& Education &10.8&16.0&18.2&20.3 \\ 
          \midrule
          \textit{N/A Decision authority} &8.5&10.9&18.4&17.1 \\
          Creation &8.5&10.9&18.4&17.1 \\
          \bottomrule
                
    \end{tabular}

\end{table}
\normalsize

\begin{table*}[]
    \small
    \centering
    \caption{Breakdown of financial services categories within each financial intent. }
    \label{tab:fs_cat_intent_stats}
    \begin{tabular}{p{3cm}p{0.3cm}p{0.3cm}p{0.3cm}p{0.3cm}p{0.3cm}p{0.3cm}p{0.3cm}p{0.3cm}p{0.3cm}p{0.3cm}p{0.3cm}p{0.3cm}p{0.3cm}p{0.3cm}p{0.3cm}p{0.3cm}p{0.3cm}p{0.3cm}}
    \toprule
         \textbf{Intent} & \multicolumn{2}{c}{\textbf{Investments}} & \multicolumn{2}{c}{\textbf{Retail}}  & \multicolumn{2}{c}{\textbf{Tax}} & \multicolumn{2}{c}{\textbf{Payments}} & \multicolumn{2}{c}{\textbf{Benefits}} & 
         \multicolumn{2}{c}{\textbf{Insurance}} & \multicolumn{2}{c}{\textbf{Business}} & 
         \multicolumn{2}{c}{\textbf{Budgeting}} & \multicolumn{2}{c}{\textbf{Finance}}\\

          \textbf{} & \multicolumn{2}{c}{\textbf{}} & \multicolumn{2}{c}{\textbf{Banking}}  & \multicolumn{2}{c}{\textbf{}} & \multicolumn{2}{c}{\textbf{\& Transfers}} & \multicolumn{2}{c}{\textbf{Public}} & 
         \multicolumn{2}{c}{\textbf{}} & \multicolumn{2}{c}{\textbf{Finance}} & 
         \multicolumn{2}{c}{\textbf{Budgeting}} & \multicolumn{2}{c}{\textbf{Infra-}}\\

         \textbf{} & \multicolumn{2}{c}{\textbf{}} & \multicolumn{2}{c}{\textbf{\& Credit}}  & \multicolumn{2}{c}{\textbf{}} & \multicolumn{2}{c}{\textbf{}} & \multicolumn{2}{c}{\textbf{Aid}} & 
         \multicolumn{2}{c}{\textbf{}} & \multicolumn{2}{c}{\textbf{}} & 
         \multicolumn{2}{c}{\textbf{}} & \multicolumn{2}{c}{\textbf{structure}}\\
\midrule
      & US & India & US & India & US & India & US & India & US & India & US & India & US & India & US & India & US & India\\
\cline{2-19}
Delegated Fin. Dec. Ex & 0.0 & 0.0 &0.0 & 0.0 & 0.0 & 0.0& 0.0 & 0.0 & 0.0 & 0.0 & 0.0 & 0.0 & 0.0 & 0.0 & 0.0 & 0.0 & 0.0 & 0.0\\
Instruction-led Fin. Ex. & 2.4 & 14.3 & 14.3 & 57.1 & 0.0 & 0.0 & 61.9 & 35.7 & 9.5 & 7.1 & 0.0 & 0.0 & 2.4 & 0.0 & 31.0 & 7.1 & 0.0 & 0.0\\
Fin. Auto. \& Monitoring  & 0.0 & 100.0 & 16.7 & 0.0 & 0.0 & 0.0 & 33.3 & 0.0 & 16.7 & 0.0 & 0.0 & 0.0 & 0.0 & 0.0 & 66.7 & 0.0 & 0.0 & 0.0\\
Prod. \& Strat. Optimisation  & 41.6 & 55.1 & 32.5 & 26.2 & 6.5 & 9.3 & 24.7 & 30.8 & 6.4 & 1.5 & 7.7 & 4.0 & 5.6 & 4.0 & 3.3 & 1.7 & 1.5 & 6.2\\
Prod. \& Strat. Comparison  &  45.0 & 57.3 & 31.9 & 26.6 & 8.2 & 12.3 & 18.9 & 25.5 & 7.2 & 1.2 & 8.5 & 4.6 & 5.4 & 4.0 & 3.1 & 1.7 & 1.9 & 5.9\\
Fin. Problem Resolution  & 14.7 & 16.7 & 37.0 & 42.6 & 9.5 & 16.7 & 30.0 & 51.9 & 21.0 & 4.6 & 6.5 & 2.7 & 3.7 & 2.1 & 3.0 & 0.7 & 6.4 & 17.0\\
Personal Fin. Analysis  & 34.9 & 57.0 & 30.6 & 27.8 & 16.6 & 18.3 & 11.5 & 12.5 & 12.7 & 2.2 & 6.7 & 3.9 & 7.3 & 4.9 & 6.7 & 2.9 & 1.9 & 3.5 \\
Fin. Planning  & 22.5 & 53.0 & 37.8 & 24.2 & 8.2 & 12.0 & 22.9 & 28.5 & 13.4 & 3.3 & 8.4 & 3.0 & 9.9 & 7.7 & 14.1 & 6.5 & 2.4 & 5.5\\
Complex Research  & 23.8 & 34.8 & 33.0 & 29.5 & 7.7 & 12.7 & 30.0 & 42.5 & 16.4 & 3.6 & 8.9 & 4.3 & 5.7 & 4.0 & 2.2 & 1.1 & 4.2 & 10.7\\
Simple Retrieval  &20.2 & 26.0 & 36.9 & 34.7 & 12.6 & 17.2 & 30.5 & 39.6 & 13.5 & 3.9 & 7.1 & 3.9 & 4.5 & 3.2 & 1.5 & 0.7 & 4.5 & 10.8\\
Fin. Learning \& Edu.  &42.0 & 48.1 & 28.1 & 24.8 & 15.7 & 21.2 & 7.0 & 12.9 & 4.9 & 3.2 & 8.9 & 6.5 & 18.0 & 14.0 & 2.5 & 3.8 & 4.3 & 9.5\\
Creation  & 22.8 & 34.7 & 27.9 & 32.2 & 11.1 & 13.7 & 22.1 & 27.1 & 10.4 & 2.9 & 10.5 & 6.4 & 12.7 & 10.3 & 8.0 & 5.4 & 6.6 & 8.6\\
\bottomrule    
    \end{tabular}
\end{table*}

\normalsize

The distribution of decision authority shows that consumers overwhelmingly use AI to inform and shape financial decisions rather than execute them (Table \ref{tab:fs_all_intent_stats}). Level 1 (\textit{Inform}) interactions form the majority of FS conversations (63.5\% of US and 72.1\% of Indian FS subchats). Within \textit{Inform} interactions, Simple Retrieval dominates both markets, indicating that LLMs frequently substitute for traditional search. Level 2 (\textit{Shape}) interactions account for 58.6\% of US and 49.7\% of India FS subchats. \textit{Shape} interactions are characterised by Financial Problem Resolution, Product \& Strategy Comparison, Product \& Strategy Optimisation and Personal Financial Analysis, showing users increasingly rely on AI to support financial judgement. By contrast, Level 3 (\textit{Act}) interactions remain infrequent, accounting for 0.3\% of US and 0.1\% of Indian FS chats. These are almost all instruction-led budgeting and tracking tasks, with close to no evidence of users delegating autonomous financial decisions to AI. This suggests that, despite interest in agentic AI, use concentrated on decision support rather than decision delegation.

Decision authority is not uniformly distributed across financial domains. Rather, different contexts exhibit distinct behavioural profiles, with optimisation concentrated in \emph{Investments} and problem resolution centred on \emph{Retail Banking \& Payments} (Table \ref{tab:fs_cat_intent_stats}).

We further examine the intents responsible for the greatest volume of high-authority interactions. Financial Problem Resolution is the largest \emph{Shape}-level intent across both markets but is substantially more prevalent in the US. Across both, these conversations concentrate on \emph{Retail Banking \& Credit} and \emph{Payments \& Transfers}, reflecting users seeking assistance with payment failures, account access, fraud, and transaction issues. In the US, a notable minority of these conversations also concern \emph{Benefits \& Public Aid}, particularly housing support, education finance and public assistance.

Product \& Strategy Optimisation is the second major \emph{Shape} behaviour. Almost one-fifth of users employ LLMs to optimise products or strategies (22.6\% and 19.3\% in US and India). These conversations are dominated by investment decisions, particularly portfolio optimisation, company analysis and investment risk assessment, with investment-related optimisation more often in India than US.

Though the overall distribution of decision authority is similar across markets, behavioural differences emerge within intent categories. Indian users make greater use of LLMs for Financial Learning \& Education, whereas US users rely more heavily on these for Financial Problem Resolution and Personal Financial Analysis.

Taken together, these findings suggest current LLMs function primarily as a decision-support technology, augmenting financial judgement while leaving ultimate decision authority with users.

\subsection{Topic Modelling}
\noindent\textit{What concrete financial activities do these behaviours correspond to?}

\begin{table}[h]
%\footnotesize
\small
\centering
\caption{Top prevalent topics and descriptive market differences. \% denote the share of FS subchats within each market.}
\label{tab:topic_prevalence}

\begin{tabularx}{\columnwidth}{@{}Xrr@{}}
\toprule
\textbf{Description} & \textbf{US (\%)} & \textbf{India (\%)} \\
\midrule
\multicolumn{3}{@{}l@{}}{\textit{Shared high-prevalence topics}}\\
Investment \& portfolio guidance                & 4.6 & 7.8 \\
Cash access, account services \& fees           & 4.1 & 5.2 \\
Transactions \& cash-flow management            & 3.9 & 4.7 \\
Payment-platform troubleshooting                & 3.5 & 3.4 \\
Tax filing, deductions \& liabilities           & 3.0 & 3.0 \\
Credit-building \& payment strategies           & 2.8 & 3.0 \\
\midrule
\multicolumn{3}{@{}l@{}}{\textit{Comparatively prevalent in the US}}\\
Insurance coverage, claims \& policy comparison & 3.5 & 1.6 \\
Education finance \& financial aid              & 2.9 & 0.5 \\
Unemployment, Social Security \& food assistance& 2.6 & 0.0 \\
\midrule
\multicolumn{3}{@{}l@{}}{\textit{Comparatively prevalent in India}}\\
Stock analysis, trading \& market research      & 1.7 & 6.6 \\
Gift-card purchase, redemption \& resale        & 0.6 & 3.5 \\
Card security, fraud \& unauthorised charges    & 0.4 & 3.4 \\
Banking products for minors                     & 0.7 & 2.8 \\
\bottomrule
\end{tabularx}
\end{table}

% The topic model identified 179 topics, of which 170 were retained after excluding nine highly user-specific topics. Table~\ref{tab:topic_prevalence} reports the principal shared topics and the largest descriptive market differences. Six of the ten most prevalent topics were common to both markets, centred on investment guidance, banking services, transaction management, payment troubleshooting, taxes, and credit-building. However, clear differences also emerged. US conversations were comparatively more focused on insurance, education finance, and public assistance, whereas Indian conversations placed greater emphasis on investment research, stock analysis, account security, gift-card transactions, and banking products for minors. These findings complement the category-level analysis by revealing the specific financial needs underlying the broader thematic differences between the two markets. As before, these comparisons are descriptive of the observed samples and should not be interpreted as population-level estimates.

Topic modelling illustrates the concrete financial activities underlying the behavioural framework. This identified 170 topics – the top 10 topics are shown in Table ~\ref{tab:topic_prevalence}, highlighting the overlap and differences in US and Indian discussions. 

Six of the ten most prevalent topics are shared across both countries, including investment guidance, payment management, banking services, taxation and credit improvement. However, important differences emerge. US conversations are more frequently concerned with insurance, education finance and public assistance, whereas Indian conversations focus more heavily on investment research, stock analysis and account security.

Rather than constituting a separate analytical framework, these topics provide concrete examples of the financial decisions and behaviours identified through the financial domain, behavioural intent and decision authority classifications.

\section{Discussion and Policy Implications}

Our findings have implications beyond characterising contemporary patterns of conversational AI use. The predominance of Level 2 (\textit{Shape}) interactions indicates that AI is already influencing consumer financial judgement at scale while rarely executing financial decisions. This intermediate behavioural space may not be fully addressed by existing consumer protection frameworks; regulators should evaluate not only whether AI executes decisions, but also the extent to which it shapes them.

For financial institutions, consumer demand is currently concentrated less on fully autonomous financial agents than on AI systems that augment human judgement. Product optimisation, financial problem resolution and personalised analysis are substantially more prevalent than execution-oriented requests, suggesting that near-term value lies in AI systems that function as cognitive decision-support tools.

Finally, the behavioural framework introduced here offers a baseline against which future AI decision making can be measured. As increasingly agentic AI systems assume greater responsibility for financial decisions, behavioural intent and delegated decision authority form a systematic framework for tracking how the role of AI shifts over time.

\section{Conclusions}

This paper introduces a measurement framework of how consumers use conversational AI in financial decision making. Rather than classifying conversations solely by topic, the framework combines behavioural intent (what users seek to accomplish) with decision authority (the extent to which judgement is offloaded to AI). Applied to 3 months of ChatGPT and Gemini chat histories from US and Indian users, we find financial services have become a substantial human-AI interaction type with consumers overwhelmingly leveraging LLMs to inform and shape finances. This framework builds a foundation for researchers, regulators and institutions to monitor the evolving role of LLMs in consumer finance.

% \section{Acknowledgments}

% Identification of funding sources and other support, and thanks to
% individuals and groups that assisted in the research and the
% preparation of the work should be included in an acknowledgment
% section, which is placed just before the reference section in your
% document.

% This section has a special environment:
% \begin{verbatim}
%   \begin{acks}
%   ...
%   \end{acks}
% \end{verbatim}
% so that the information contained therein can be more easily collected
% during the article metadata extraction phase, and to ensure
% consistency in the spelling of the section heading.

% Authors should not prepare this section as a numbered or unnumbered {\verb|\section|}; please use the ``{\verb|acks|}'' environment.

%%
%% The acknowledgments section is defined using the "acks" environment
%% (and NOT an unnumbered section). This ensures the proper
%% identification of the section in the article metadata, and the
%% consistent spelling of the heading.
% \begin{acks}
% To Robert, for the bagels and explaining CMYK and color spaces.
% \end{acks}

\section*{Ethics and Privacy Statement}

This research analyses voluntarily contributed and reimbursed human–AI conversations collected by a third-party data provider under informed participant consent. Analysis used de-identified data, and results are reported only in aggregate without identifying individuals. By characterising how consumers allocate decision authority to AI, the framework and models introduced here provide a basis for informing the development of safer, more transparent, and more accountable AI systems for financial decision making.

%This section of your ACM work should discuss the potential societal risks that might result from its publication; two to three sentences related to the findings of your study, or new advancements made possible by their developed methods. The privacy and ethics statement should clearly address the broader impacts of their work as it relates to the authors' interpretation of privacy, fairness, safety, human rights, data sovereignty, or future misuse and any benefit/risk trade-off resulting from this research. We acknowledge that some papers may have minimal societal risks beyond those considered by institutional review boards, and the dimensions considered by any review of the user study design or dataset licenses could be provided in this statement.

%%
%% The next two lines define the bibliography style to be used, and
%% the bibliography file.
\bibliographystyle{ACM-Reference-Format}
\bibliography{sample-base}

@techreport{chatterji2025how,
  title        = {How People Use ChatGPT},
  author       = {Chatterji, Aaron and Cunningham, Tom and Deming, David J. and Hitzig, Zo{\"e} and Ong, Christopher and Shan, Carl Yan and Wadman, Kevin},
  institution  = {National Bureau of Economic Research},
  number       = {34255},
  year         = {2025},
  type         = {NBER Working Paper},
  url          = {https://www.nber.org/papers/w34255}
}

@misc{handa2025anthropic,
  title        = {The Anthropic Economic Index},
  author       = {Handa, Anmol and others},
  year         = {2025},
  howpublished = {Anthropic Research Report},
  url          = {https://www.anthropic.com/news/the-anthropic-economic-index}
}

@article{costagomes2025time,
  title   = {It's About Time: The Temporal and Modal Dynamics of Copilot Usage},
  author  = {Costa-Gomes, Beatriz and Chen, Sophia and Hsueh, Connie and Morgan, Deborah and Schoenegger, Philipp and Shah, Yash and Way, Samuel and Zhu, Yuki and Adeline, Timoth{\'e} and Bhaskar, Michael and Suleyman, Mustafa and Spielman, Seth},
  journal = {arXiv preprint arXiv:2512.11879},
  year    = {2025},
  url     = {https://arxiv.org/abs/2512.11879}
}

@article{costagomes2026health,
  author = {Costa-Gomes, Beatriz and Tolmachev, Pavel and Taysom, Eloise and Sounderajah, Viknesh and Richardson, Hannah and Schoenegger, Philipp and Liu, Xiaoxuan and Nour, Matthew M. and Spielman, Seth and Way, Samuel F. and Shah, Yash and Bhaskar, Michael and Nori, Harsha and Kelly, Christopher and Hames, Peter and Gross, Bay and Suleyman, Mustafa and King, Dominic},
  title = {Public use of a generalist LLM chatbot for health queries},
  journal = {Nature Health},
  year = {2026},
  volume = {1},
  number = {7},
  pages = {689--696},
  doi = {10.1038/s44360-026-00117-x},
  url = {https://doi.org/10.1038/s44360-026-00117-x},
  issn = {3005-0693}
}

@article{roychowdhury2026conversation,
  title   = {How People Use ChatGPT: Conversation-Level Evidence from India, Nigeria, Brazil and Pakistan},
  author  = {Roy Chowdhury, G. V. and Garimella, Kiran},
  year    = {2026},
  note    = {Preprint},
  url     = {https://gvrkiran.github.io/content/How_people_use_ChatGPT.pdf}
}

@article{paydarzarnaghi2026financechatgpt,
  title={What Do People Ask AI About Finance? Evidence from ChatGPT},
  author={Paydarzarnaghi, M. and others},
  year={2026},
  journal={SSRN Electronic Journal},
  url={https://papers.ssrn.com/sol3/papers.cfm?abstract_id=6999350}
}

@article{pak2026personalfinance,
  title={How Individuals Use Generative AI for Personal Financial Management},
  author={Pak, ...},
  journal={International Journal of Information Management Data Insights},
  year={2026}
}

@article{choukhmane2025financialadvice,
  title={AI Financial Advice: Supply, Demand, and Life-Cycle Effects},
  author={Choukhmane, Taha and others},
  year={2025},
  note={Working paper}
}

@article{niszczota2023financially,
  title={GPT Has Become Financially Literate: Insights from Financial Literacy Tests of GPT and a Preliminary Test of How People Use It as a Source of Advice},
  author={Niszczota, Pawe{\l} and Abbas, Sami},
  journal={arXiv preprint arXiv:2309.00649},
  year={2023}
}

@misc{lloyds2025digitalindex,
  title={Consumer Digital Index 2025},
  author={{Lloyds Banking Group}},
  year={2025},
  howpublished={\url{https://www.lloydsbankinggroup.com/assets/pdfs/media/consumer-digital-index/2025/2025-consumer-digital-index.pdf}}
}

@inproceedings{casanueva2020banking77,
  title={Efficient Intent Detection with Dual Sentence Encoders},
  author={Casanueva, I{\~n}igo and Tem{\v c}inas, Pavel and Gerz, Daniela and Henderson, Matthew and Vuli{\'c}, Ivan},
  booktitle={Proceedings of the 2nd Workshop on NLP for Conversational AI},
  pages={38--45},
  year={2020}
}

@article{Beltagy2020Longformer,
  title={Longformer: The Long-Document Transformer},
  author={Iz Beltagy and Matthew E. Peters and Arman Cohan},
  journal={arXiv:2004.05150},
  year={2020},
}

@inproceedings{vaswani2017,
author = {Vaswani, Ashish and Shazeer, Noam and Parmar, Niki and Uszkoreit, Jakob and Jones, Llion and Gomez, Aidan N. and Kaiser, \L{}ukasz and Polosukhin, Illia},
title = {Attention is all you need},
year = {2017},
isbn = {9781510860964},
publisher = {Curran Associates Inc.},
address = {Red Hook, NY, USA},
pages = {6000–6010},
numpages = {11},
location = {Long Beach, California, USA},
series = {NIPS'17}
}

@inproceedings{devlin-etal-2019-bert,
    title = "{BERT}: Pre-training of Deep Bidirectional Transformers for Language Understanding",
    author = "Devlin, Jacob  and
      Chang, Ming-Wei  and
      Lee, Kenton  and
      Toutanova, Kristina",
    editor = "Burstein, Jill  and
      Doran, Christy  and
      Solorio, Thamar",
    booktitle = "Proceedings of the 2019 Conference of the North {A}merican Chapter of the Association for Computational Linguistics: Human Language Technologies, Volume 1 (Long and Short Papers)",
    month = jun,
    year = "2019",
    address = "Minneapolis, Minnesota",
    publisher = "Association for Computational Linguistics",
    url = "https://aclanthology.org/N19-1423/",
    doi = "10.18653/v1/N19-1423",
    pages = "4171--4186",
}

@article{shah2025,
author = {Shah, Chirag and White, Ryen and Andersen, Reid and Buscher, Georg and Counts, Scott and Das, Sarkar and Montazer, Ali and Manivannan, Sathish and Neville, Jennifer and Rangan, Nagu and Safavi, Tara and Suri, Siddharth and Wan, Mengting and Wang, Leijie and Yang, Longqi},
title = {Using Large Language Models to Generate, Validate, and Apply User Intent Taxonomies},
year = {2025},
issue_date = {August 2025},
publisher = {Association for Computing Machinery},
address = {New York, NY, USA},
volume = {19},
number = {3},
issn = {1559-1131},
url = {https://doi.org/10.1145/3732294},
doi = {10.1145/3732294},
journal = {ACM Trans. Web},
month = aug,
articleno = {34},
numpages = {29}
}

@misc{shelby2025taxonomyuserneedsactions,
      title={Taxonomy of User Needs and Actions}, 
      author={Renee Shelby and Fernando Diaz and Vinodkumar Prabhakaran},
      year={2025},
      eprint={2510.06124},
      archivePrefix={arXiv},
      primaryClass={cs.HC},
      url={https://arxiv.org/abs/2510.06124}, 
}

@inproceedings{bigbird,
author = {Zaheer, Manzil and Guruganesh, Guru and Dubey, Avinava and Ainslie, Joshua and Alberti, Chris and Ontanon, Santiago and Pham, Philip and Ravula, Anirudh and Wang, Qifan and Yang, Li and Ahmed, Amr},
title = {Big bird: transformers for longer sequences},
year = {2020},
isbn = {9781713829546},
publisher = {Curran Associates Inc.},
address = {Red Hook, NY, USA},
articleno = {1450},
numpages = {15},
location = {Vancouver, BC, Canada},
series = {NIPS '20}
}

@article{WARKULAT2024103721,
title = {Social media attention and retail investor behavior: Evidence from r/wallstreetbets},
journal = {International Review of Financial Analysis},
volume = {96},
pages = {103721},
year = {2024},
issn = {1057-5219},
doi = {https://doi.org/10.1016/j.irfa.2024.103721},
url = {https://www.sciencedirect.com/science/article/pii/S1057521924006537},
author = {Sonja Warkulat and Matthias Pelster}
}

@article{cao2020,
author = {Cao, Yingxia and Gong, Fengmei and Zeng, Tong},
year = {2020},
month = {03},
pages = {JFCP-18},
title = {Antecedents and Consequences of Using Social Media for Personal Finance},
volume = {31},
journal = {Journal of Financial Counseling and Planning},
doi = {10.1891/JFCP-18-00049}
}

@inproceedings{theerthala-2025-synthesizing,
    title = "Synthesizing Behaviorally-Grounded Reasoning Chains: A Data-Generation Framework for Personal Finance {LLM}s",
    author = "Theerthala, Akhil",
    editor = "Chen, Chung-Chi  and
      Winata, Genta Indra  and
      Rawls, Stephen  and
      Das, Anirban  and
      Chen, Hsin-Hsi  and
      Takamura, Hiroya",
    booktitle = "Proceedings of The 10th Workshop on Financial Technology and Natural Language Processing",
    month = nov,
    year = "2025",
    address = "Suzhou, China",
    publisher = "Association for Computational Linguistics",
    url = "https://aclanthology.org/2025.finnlp-2.11/",
    doi = "10.18653/v1/2025.finnlp-2.11",
    pages = "167--190"
}

@inproceedings{lu-huo-2025-financial,
    title = "Financial Named Entity Recognition: How Far Can {LLM} Go?",
    author = "Lu, Yi-Te  and
      Huo, Yintong",
    editor = "Chen, Chung-Chi  and
      Moreno-Sandoval, Antonio  and
      Huang, Jimin  and
      Xie, Qianqian  and
      Ananiadou, Sophia  and
      Chen, Hsin-Hsi",
    booktitle = "Proceedings of the Joint Workshop of the 9th Financial Technology and Natural Language Processing (FinNLP), the 6th Financial Narrative Processing (FNP), and the 1st Workshop on Large Language Models for Finance and Legal (LLMFinLegal)",
    month = jan,
    year = "2025",
    address = "Abu Dhabi, UAE",
    publisher = "Association for Computational Linguistics",
    url = "https://aclanthology.org/2025.finnlp-1.15/",
    pages = "164--168",
}

@article{calinski1974dendrite,
  title={A Dendrite Method for Cluster Analysis},
  author={Cali{\'n}ski, Tadeusz and Harabasz, Jerzy},
  journal={Communications in Statistics},
  volume={3},
  number={1},
  pages={1--27},
  year={1974},
  doi={10.1080/03610927408827101}
}

@inproceedings{roder2015exploring,
  title={Exploring the Space of Topic Coherence Measures},
  author={R{\"o}der, Michael and Both, Andreas and Hinneburg, Alexander},
  booktitle={Proceedings of the Eighth ACM International Conference on Web Search and Data Mining},
  pages={399--408},
  year={2015},
  doi={10.1145/2684822.2685324}
}

@article{rousseeuw1987silhouettes,
  title={Silhouettes: A Graphical Aid to the Interpretation and Validation of Cluster Analysis},
  author={Rousseeuw, Peter J.},
  journal={Journal of Computational and Applied Mathematics},
  volume={20},
  pages={53--65},
  year={1987},
  doi={10.1016/0377-0427(87)90125-7}
}

@article{grootendorst2022bertopic,
  title={BERTopic: Neural topic modeling with a class-based TF-IDF procedure},
  author={Grootendorst, Maarten},
  journal={arXiv preprint arXiv:2203.05794},
  year={2022}
}

@article{zhang2025qwen3embedding,
  title={Qwen3 Embedding: Advancing Text Embedding and Reranking Through Foundation Models},
  author={Zhang, Yanzhao and Li, Mingxin and Long, Dingkun and Zhang, Xin and Lin, Huan and Yang, Baosong and Xie, Pengjun and Yang, An and Liu, Dayiheng and Lin, Junyang and Huang, Fei and Zhou, Jingren},
  journal={arXiv preprint arXiv:2506.05176},
  year={2025}
}

@inproceedings{Joshi2020State,
  author    = {Pratik Joshi and Sebastin Santy and Amar Budhiraja and Kalika Bali and Monojit Choudhury},
  title     = {The State and Fate of Linguistic Diversity and Inclusion in the NLP World},
  booktitle = {Proceedings of the 58th Annual Meeting of the Association for Computational Linguistics},
  pages     = {6282--6293},
  year      = {2020},
  publisher = {Association for Computational Linguistics}
}

@article{Park2016TopicDrift,
  author  = {Albert Park and Andrea L. Hartzler and Jina Huh and Gary Hsieh and David W. McDonald and Wanda Pratt},
  title   = {{``How Did We Get Here?''}: Topic Drift in Online Health Discussions},
  journal = {Journal of Medical Internet Research},
  volume  = {18},
  number  = {11},
  pages   = {e284},
  year    = {2016},
  doi      = {10.2196/jmir.6297}
}

@article{Krumpal2013,
  author  = {Ivar Krumpal},
  title   = {Determinants of Social Desirability Bias in Sensitive Surveys: A Literature Review},
  journal = {Quality \& Quantity},
  volume  = {47},
  number  = {4},
  pages   = {2025--2047},
  year    = {2013},
  doi     = {10.1007/s11135-011-9640-9}
}

@inproceedings{muennighoff2022mteb,
    title = "{MTEB}: Massive Text Embedding Benchmark",
    author = "Muennighoff, Niklas  and
      Tazi, Nouamane  and
      Magne, Loic  and
      Reimers, Nils",
    editor = "Vlachos, Andreas  and
      Augenstein, Isabelle",
    booktitle = "Proceedings of the 17th Conference of the European Chapter of the Association for Computational Linguistics",
    month = may,
    year = "2023",
    address = "Dubrovnik, Croatia",
    publisher = "Association for Computational Linguistics",
    url = "https://aclanthology.org/2023.eacl-main.148/",
    doi = "10.18653/v1/2023.eacl-main.148",
    pages = "2014--2037",
}

@misc{google_ai_economy,
    author       = {{Google}},
    title        = {{AI \& Economy Research Program}},
    howpublished = {\url{https://ai.google/economy/}},
    note         = {Accessed: 2026-07-27}
  }

@article{he2023llmhealthcare,
  author       = {Kai He and Rui Mao and Qika Lin and Yucheng Ruan and Xiang Lan and Mengling Feng and Erik Cambria},
  title        = {A Survey of Large Language Models for Healthcare: From Data, Technology, and Applications to Accountability and Ethics},
  journal      = {arXiv preprint arXiv:2310.05694},
  year         = {2023},
  eprint       = {2310.05694},
  archivePrefix= {arXiv},
  primaryClass = {cs.CL}
}

@article{nie2024llmfinance,
  author       = {Wei Nie and Yuhang Li and Jiyao Wang and Yixuan Wang and Zhen Zhang and Shuai Lu},
  title        = {A Survey of Large Language Models for Financial Applications: Progress, Challenges, and Future Directions},
  journal      = {arXiv preprint arXiv:2406.11903},
  year         = {2024},
  eprint       = {2406.11903},
  archivePrefix= {arXiv},
  primaryClass = {cs.CL}
}

%%
%% If your work has an appendix, this is the place to put it.

\end{document}